\documentclass{webofc}
\usepackage[varg]{txfonts}   
\begin{document}
\title{Standard sirens and dark sector with Gaussian process~\footnote{The invited talk is given by the first author of this paper.}}
%
%

\author{ Rong-Gen  Cai~\inst{1,2}\fnsep\thanks{\email{cairg@itp.ac.cn} }, Tao Yang~\inst{3}\fnsep\thanks{\email{yangtao2017@bnu.edu.cn}}
}

\institute{CAS Key Laboratory of Theoretical Physics, Institute of Theoretical Physics,
Chinese Academy of Sciences, P.O. Box 2735, Beijing 100190, China
\and
          School of Physical Sciences, University of Chinese Academy of Sciences, No. 19A Yuquan Road, Beijing 100049, China
 \and
Department of Astronomy, Beijing Normal University, Beijing, 100875, China 
 }

\abstract{%
 The gravitational waves from compact binary systems are viewed as a standard siren to probe the evolution of the universe. 
 This paper summarizes the potential and ability to use the gravitational waves to constrain the cosmological parameters and the dark sector interaction in the Gaussian process methodology. After briefly introducing the method to reconstruct
the dark sector  interaction  by the Gaussian process, the concept of standard sirens and the analysis of reconstructing the
dark sector interaction with LISA are outlined. Furthermore, we  estimate the constraint ability of the gravitational waves on cosmological parameters
with ET. The numerical methods we use are Gaussian process and the Markov-Chain Monte-Carlo. Finally,  we also forecast
the improvements of the abilities to constrain the cosmological parameters with ET and LISA combined with the \it Planck.
}
\maketitle
\section{Introduction}
\label{intro}
More than 100 years,  Einstein himself  have predicated the existence of  the gravitational waves (GWs) by his general relativity of gravitational theory. 
 On 11 February 2016, the Laser Interferometer Gravitational Wave
Observatory (LIGO) collaboration reported the first direct detection of the gravitational wave
source GW150914~\cite{Abbott:2016blz}, which means new era of cosmology and astronomy with GW is coming.  So far another two gravitational wave 
signals have also been detected~\cite{Abbott:2016nmj,Abbott:2017vtc}. These three signals are all from the merger of two black holes. 
The information gathered from these GW sources will lead to advances not only in fundamental
physics and astrophysics, but also in cosmology~\cite{Cai:2017cbj}. In this paper, we summarize some of our  recent works
in which we use GWs as the standard sirens and explore its ability in constraining the cosmological
parameters~\cite{Cai:2015zoa,Cai:2017yww,Cai:2016sby}. Our works are inspired by Schutz's idea which was proposed in 1986~\cite{Schutz:1986gp}.
Schutz showed that it is possible to determine the Hubble constant from GW observations, by using the
fact that GWs from the binary systems encode the absolute distance information. We can assume other
techniques are available to obtain the redshift of a GW event, for example, we can measure the redshift through the identification of an accompanying electromagnetic (EM) signal which is called EM counterpart. Thus we can get the $d_L-z$ relation and use it to explore the cosmic expansion. We use the Einstein Telescope~\footnote{\url{http://www.et-gw.eu/}} (ET, the third-generation ground-based detector) and the Laser Interferometric Space
Antenna~\footnote{\url{https://www.elisascience.org/}} (LISA, the space based detector) to simulate the GWs data. ET is mainly envisaged to detect the GWs produced by the binary systems consisting of the neutron stars (NSs) and black holes (BHs). The frequency range is about $1-10^4$ Hz. We use it to forecast the GW's constraints on cosmological parameters such as the dark matter density $\Omega_m$, Hubble constant $H_0$, and the equation of state $w$ of dar energy, while, the LISA mission aims at measuring gravitational waves (GWs) in the frequency band around the mHz. Among the richness of astrophysical sources expected to produce a detectable GW signal at those frequencies, the signal produced by the massive black hole binaries (MBHBs) from $10^4$ to $10^7$ solar masses is expected to the loudest ~\cite{Tamanini:2016uin}. For the MBHB standard siren datasets constructed with LISA contain events that could be detected up to $z\sim 10$ which is much larger than that of the type Ia supernova (SNIa), we expect that GW information can extend the reconstructed dark interaction to higher redshifts. In what follows, after briefly reviewing how the dark sector interaction can be reconstructed by GP~\cite{Cai:2015zoa}, the concept of standard sirens and the analysis of reconstructing the
dark sector with LISA from~\cite{Cai:2017yww} are introduced. Then we show the results of using the ET to constrain the
cosmological parameters in~\cite{Cai:2016sby}. Finally, we give the conclusions and future prospects.

\section{Dark sector interaction with Gaussian process}
\label{sec:darksector_GP}
The cosmological constant together with the cold dark matter (CDM) (called the $\Lambda$CDM model) turned
out to be the standard model which fits the current observational data sets consistently.
However, it is faced with the fine-tuning problem and the coincidence problem. Especially, to alleviate the coincident problem,an interaction between the dark energy and dark matter is introduced.
Usually, to study the interaction of the dark energy and dark matter, the interaction form has to be assumed (for a review, see~\cite{Wang:2016lxa}). Such an assumed  interaction form
between dark energy and dark matter may introduce some bias.   Therefore a model independent investigation is called for.  In~\cite{Cai:2009ht}, we investigated the possible interaction in a way independent of specific interacting form by dividing
the  whole range of redshift into a few bins and setting the interacting term to be a constant in each redshift, it was found that the interaction is likely to cross the non-interacting
line and has an oscillation behavior. Some years later, Salvatelli {\it et al}~\cite{Salvatelli:2014zta} showed that the null interaction is excluded at $99\%$ confidence level (C.L.) when
they added the redshift-space distortions (RSD) data to the {\it Planck} data for the decaying vacuum energy model. 

\subsection{The model}
\label{sec:the_model}
In~\cite{Cai:2015zoa}, we present a nonparametric approach to reconstruct the interaction term between dark energy and dark matter directly from the observational data using the Gaussian process.
In a flat universe with an interaction between dark energy and dark matter, we can change the conservation equation to be
\begin{equation}
\dot\rho_m+3H\rho_m =  - Q\,,
\end{equation}
\begin{equation}
\dot\rho_{DE}+3H(1+w)\rho_{DE} = Q\,.
\end{equation}
Here we don't parameterize the interaction form $Q$ to be any form. Combing the Friedmann equation
and after some calculations, one can finally get
\begin{align}
- wq = &~2\left(\frac{{3D'{'^2}}}{{D{'^5}}} - \frac{{D'''}}{{D{'^4}}} + \frac{{w'D''}}{{wD{'^4}}}\right){(1 + z)^2}
          + \left[2(5 + 3w)\frac{{D''}}{{D{'^4}}} + \frac{{3w'}}{{wD{'^3}}}\right](1 + z) \nonumber\\
         &~+ \frac{{9(1 + w)}}{{D{'^3}}}\,,
\label{equa:qD}
\end{align}
where, $Q=qH_0^3$ and $D(z)=(H_0/c)(1+z)^{-1} d_L(z)$. From this, we can see that once the equation of state $w$ of dark energy is given, we can use the observed distance-redshift relationship to reconstruct the interaction.

\subsection{The Gaussian process}
The Gaussian process~\footnote{\url{http://www.gaussianprocess.org/gpml/}} is designed to use the supervised learning to build a model (or function)
from the training data and then use the model to forecast the new samples. In our works, it can
reconstruct a function from data without assuming a parametrization for it. The GP method has been used in many works~\cite{Holsclaw:2010nb,Holsclaw:2010sk,Holsclaw:2011wi}, We use the GaPP
code developed in~\cite{Seikel:2012uu} to derive our results. The distribution over functions  provided by GP is suitable to describe the observed data. At each point $z$, the reconstructed function $f(z)$ is also a Gaussian distribution with a mean value and Gaussian error. The functions at different points $z$ and $\tilde{z}$ are related by a covariance function $k(z,\tilde{z})$, which only depends on a set of hyperparameters $\ell$ and $\sigma_f$. Here $\ell$ gives a measure of the coherence length of the correlation in $x$-direction and $\sigma_f$ denotes the overall amplitude of the correlation in the $y$-direction. Both of them will be optimized by GP with the observed data set.
In contrast to actual parameters, GP does not specify the form of the reconstructed function. Instead it characterizes the typical changes of the function. According to~\cite{Seikel:2013fda}, we choose the kernel function as the mat\'{e}rn ($\nu=9/2$) form
\begin{align}
k(z,\tilde z) = {\sigma _f}^2\exp\left( - \frac{{3\left| {z - \tilde z} \right|}}{\ell }\right) \times\Big[1 + \frac{{3\left| {z - \tilde z} \right|}}{\ell } + \frac{{27{{(z - \tilde z)}^2}}}{{7{\ell ^2}}}+ \frac{{18{{\left| {z - \tilde z} \right|}^3}}}{{7{\ell ^3}}} + \frac{{27{{(z - \tilde z)}^4}}}{{35{\ell ^4}}}\Big]\,.
\end{align}
For detailed description of the Gaussian process, one can refer to~\cite{Seikel:2012uu}, see also \cite{Cai:2015zoa, Cai:2015pia,Cai:2016vmn}.

\subsection{Reconstruct the dark sector interaction}
\begin{figure}[h]
\centering
\includegraphics[width=0.3\textwidth]{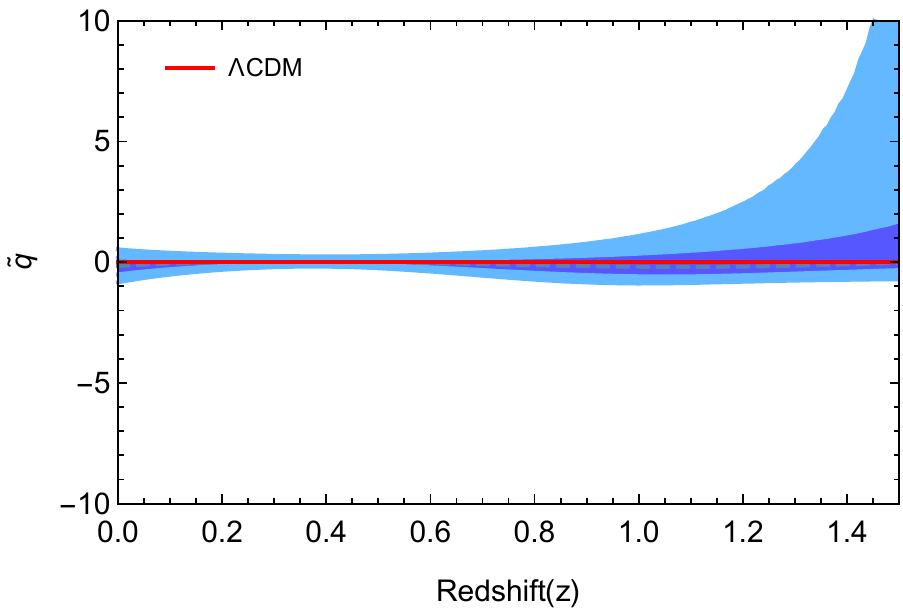}
\includegraphics[width=0.3\textwidth]{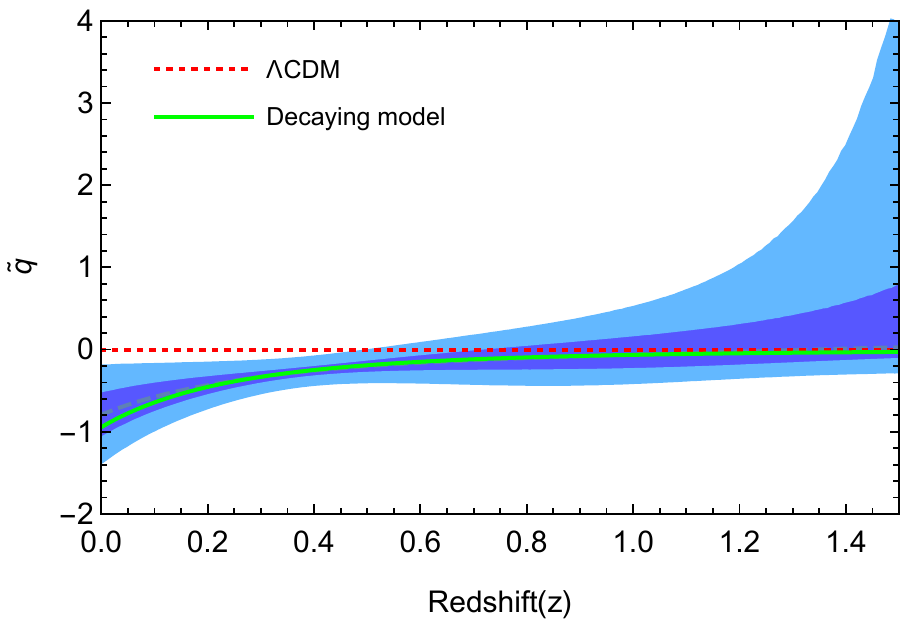}
\includegraphics[width=0.3\textwidth]{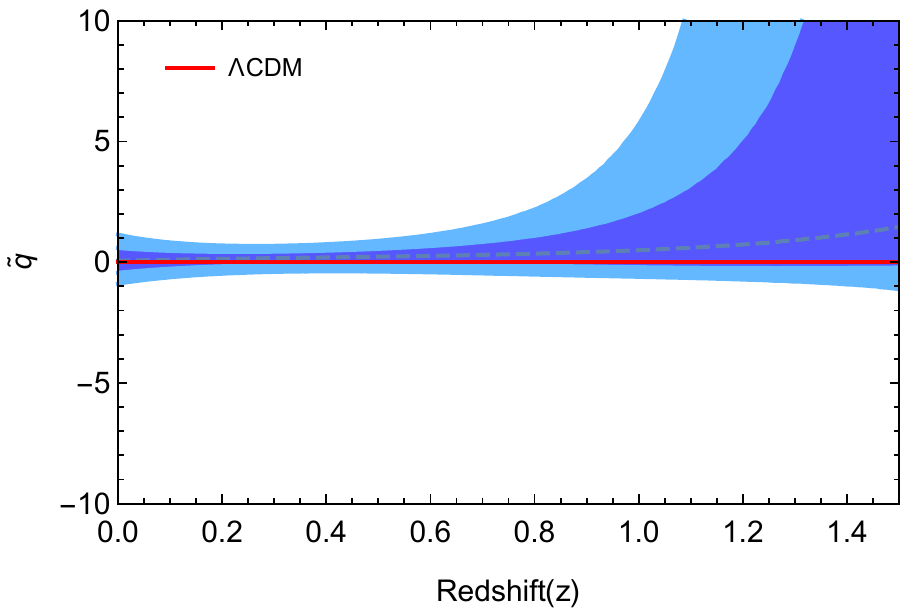}
\caption{The results from~\cite{Cai:2015zoa} for the reconstruction of the interaction term. The left is from the mock data of DES for the $\Lambda$CDM model, the middle is from the mock data of DES for the decaying model, and the right is from the real data sets Union2.1 for the
vacuum dark energy and dark matter. Here $\tilde q(z) = q /(1+z)^6$.}
\label{fig:dark_GP}       
\end{figure}

To see the ability of the GP method to distinguish different models and recover the correct behaviors of the models, we create mock data sets of future  SNIa according to Dark Energy Survey (DES)~\cite{Bernstein:2011zf} for two fiducial models: the LCDM model and a decaying vacuum energy model
with $\rho_{DE}=3\alpha H$ with $w=-1$. Here we set $\Omega_{m0}=0.3$ for both. The mock data results are shown in figure~\ref{fig:dark_GP} (left and middle). We can see the GP method can recover and distinguish both of these two fiducial models. Now we can apply our GP method to the real data
set Union2.1. We firstly test the vacuum dark energy with the dark matter. The result is shown in figure~\ref{fig:dark_GP} (right). It tells us that the $\Lambda$CDM model (no interaction) fits well with the SNIa Union 2.1 data sets. More results for the $w$CDM model and CPL parametrization can be found in~\cite{Cai:2015zoa}.

Here we have reviewed how to reconstruct the dark sector interaction by Gaussian process. In the following, we will introduce our works on using the GWs as the standard sirens to reconstruct the dark sector interaction~\cite{Cai:2017yww} and constrain the cosmological parameters~\cite{Cai:2016sby}.

\section{Reconstruct the dark sector with LISA}
In this section, we summarize the  results from the paper~\cite{Cai:2017yww} in which we use the MBHBs as the standard sirens to reconstruct the dark sector. In the most recent configuration for LISA, conceived to answer the call from ESA for its L3 mission, the observatory will be based on three arms with six active laser links, between three identical spacecraft in a triangular formation separated by 2.5 million km. The proposed nominal mission duration in science mode is of 4 years, but the mission should be designed with consumables and orbital stability to facilitate a total mission duration of up to 10 years.
The simulated datasets of MBHB standard sirens are taken from~\cite{Tamanini:2016zlh}, where realistic catalogues of GW events detected by
LISA and for which an EM counterpart can be identified by future telescopes were constructed.
The MBHB standard siren datasets constructed with LISA contain events that could be detected up to $z\sim10$. It is
thus expected that GW information can extend the reconstructed dark interaction to a much higher redshift.

\subsection{Massive black hole binary mergers as standard sirens for LISA}
Compact binary inspirals are excellent examples of standard sirens, irrespective of the masses of the inspiralling objects.
In fact the waveform $h$ produced by these systems during the inspiral phase, is theoretically well described by the analytical solution
\begin{align}
	h_\times = \frac{4}{d_L(z)} \left( \frac{G \mathcal{M}_c(z)}{c^2} \right)^{5/3} \left( \frac{\pi f}{c} \right)^{2/3} \cos\iota \sin[\Phi(f)] \,,
	\label{eq:gwf}
\end{align}
which is valid at the lowest (Newtonian) order for the ``cross'' GW polarization (the same expression holds for the ``plus'' polarization with a difference dependence on the orientation of the binary's orbital plane $\iota$).
Here $\mathcal{M}_c(z)$ is the (redshifted) chirp mass, $f$ the GW frequency at the observer and $\Phi(f)$ its phase.
For our purposes the luminosity distance $d_L$(z) is the most important parameter entering the waveform. Parameter estimation
over the observed GW signal directly yields the value of the luminosity distance of the GW source, together with an uncertainty due to the detector noise. This implies that once the signal is detected and characterised, the luminosity distance of the source can be extracted, without the need of cross calibrating with other distance indicators or unwanted systematics, as in the case of SNIa. If also the redshift $z$ of the GW source (or the one of its
hosting galaxy) is measured, one obtains a data point in the distant-redshift diagram which can be used to constrain the distance-redshift relation.

The point now is to predict how many MBHB standard sirens will be detected by LISA and what kind of cosmological data they will provide. Based on
~\cite{Barausse:2012fy}, Three different competing scenarios for the initial conditions of the massive BH population at high redshift have been
taken into account:
\begin{itemize}
	\item {\bf popIII}: a ``light-seeds'' scenario where the massive BHs form from the remnants of population III stars;
	\item {\bf Q3d}: a ``heavy-seeds'' scenario where massive BHs form from the collapse of protogalactic disks, but a time delay between the merger of the hosting galaxies and the BHs is present;
	\item {\bf Q3nod}: another ``heavy-seeds'' scenario with no time delay between the merger of hosting galaxies and BHs.
\end{itemize}
There are 118 catalogues of 5 years produced for each MBHB formation model to reduce the statistical uncertainty due to the low number of events in each catalogue. We select a representative catalogue among all 118 of them by the figure of merit. This will allow us to use the Gaussian process
using the representative catalogue, which provides a fairly realistic realisation of MBHB standard sirens data detected by LISA.  We will also repeat the analysis considering a LISA mission duration of 10 years, for which the number of standard sirens events on average doubles with respect to a 5 years mission.

\subsection{The results}
\begin{figure}[h]
\centering
\includegraphics[width=0.3\textwidth]{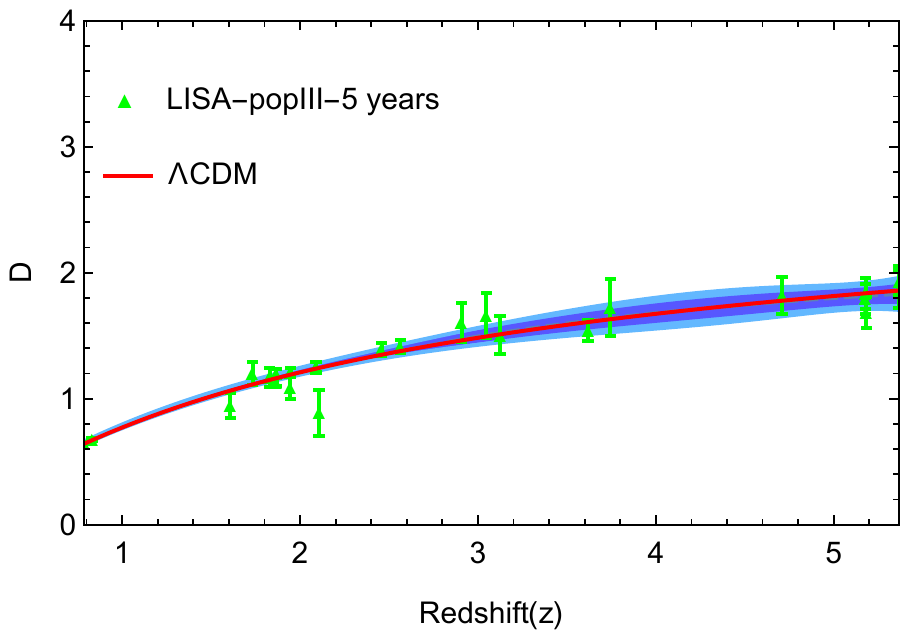}
\includegraphics[width=0.3\textwidth]{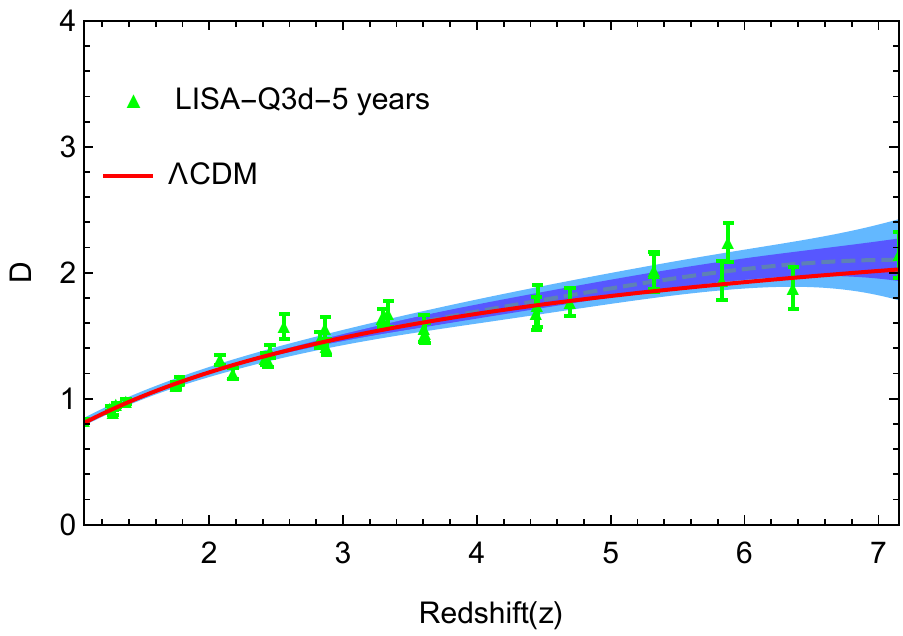}
\includegraphics[width=0.3\textwidth]{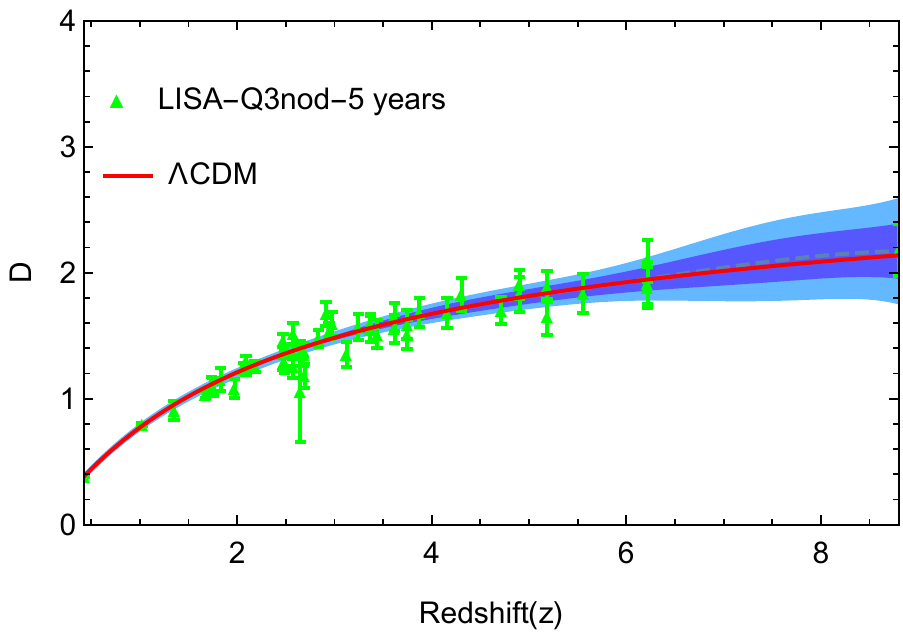}\\
\includegraphics[width=0.3\textwidth]{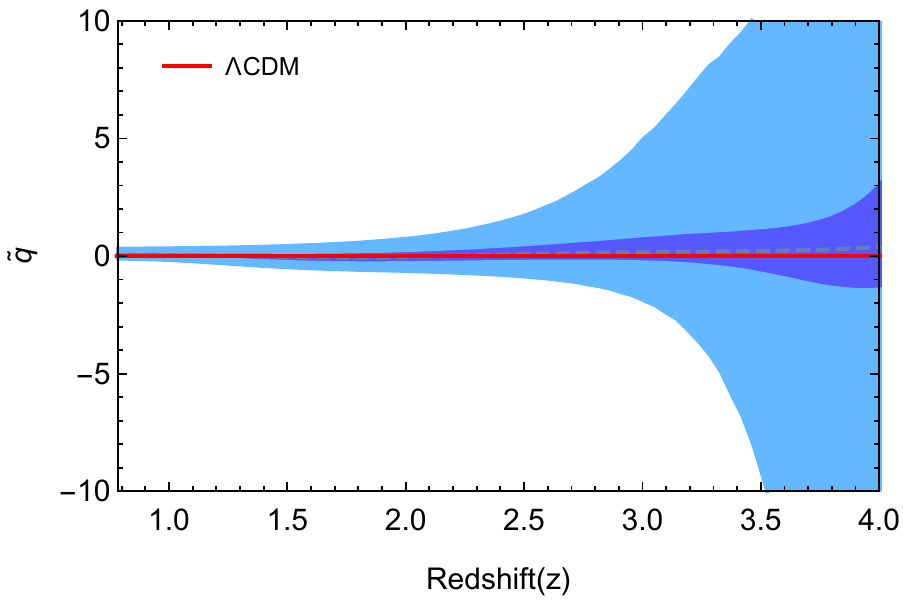}
\includegraphics[width=0.3\textwidth]{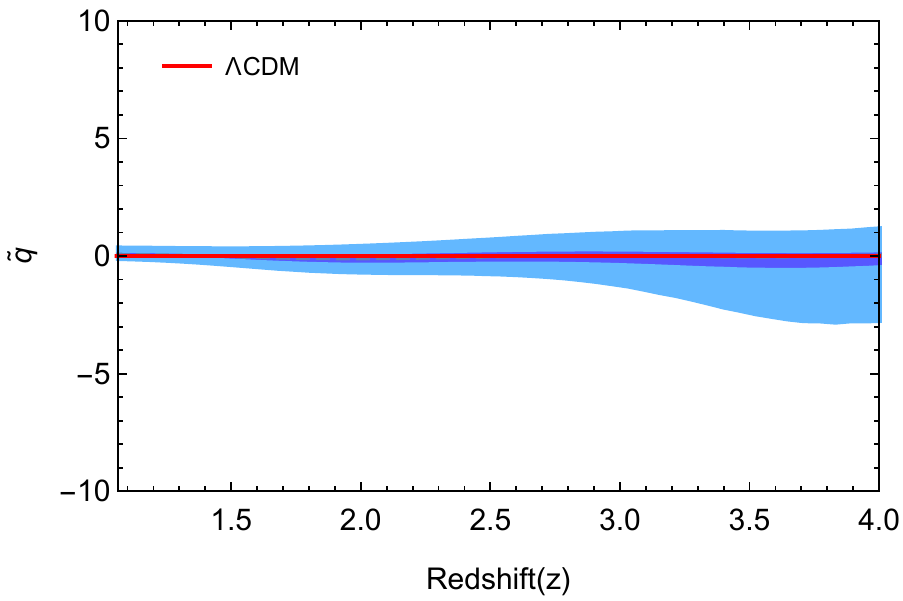}
\includegraphics[width=0.3\textwidth]{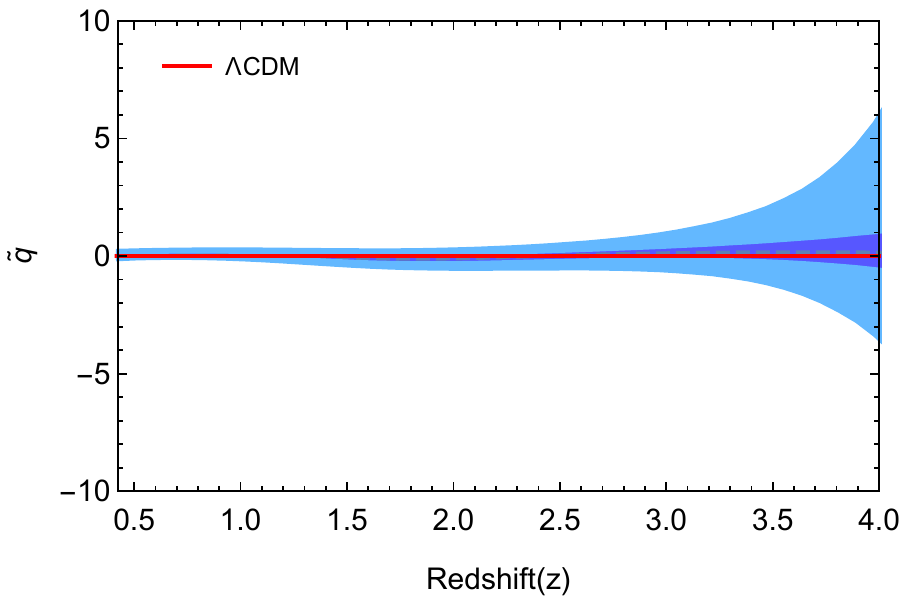}
\caption{Reconstructions of the interaction using LISA alone for a 5 years mission. From left
  to right each column reports the results for popIII, Q3d, Q3nod.}
\label{fig:dark_LISA_5y}       
\end{figure}

\begin{figure}[h]
\centering
\includegraphics[width=0.3\textwidth]{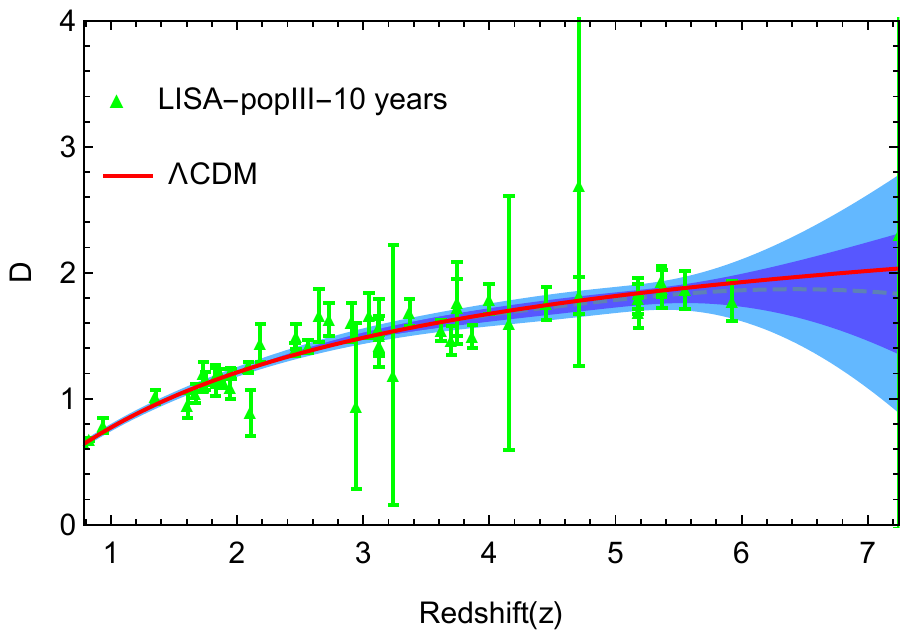}
\includegraphics[width=0.3\textwidth]{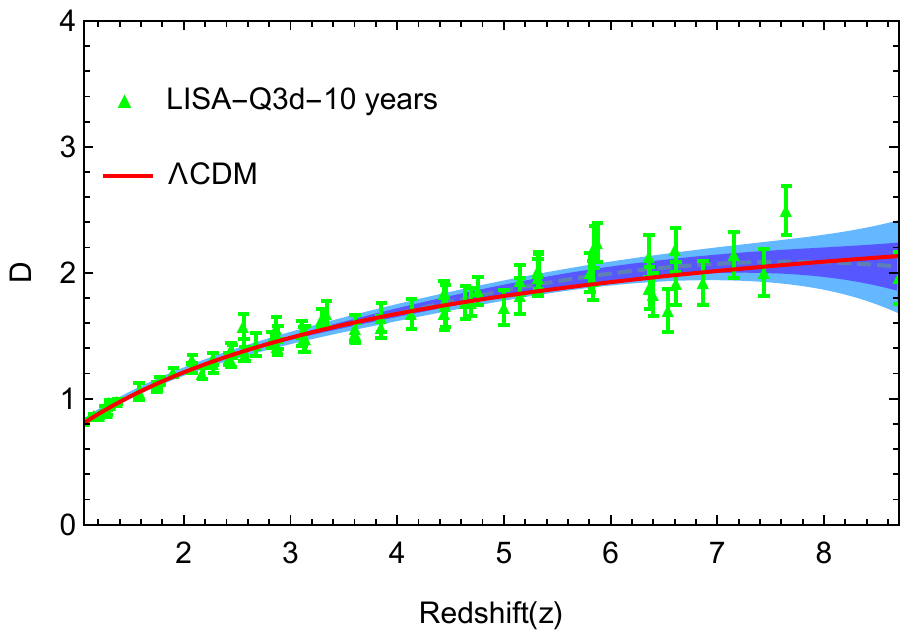}
\includegraphics[width=0.3\textwidth]{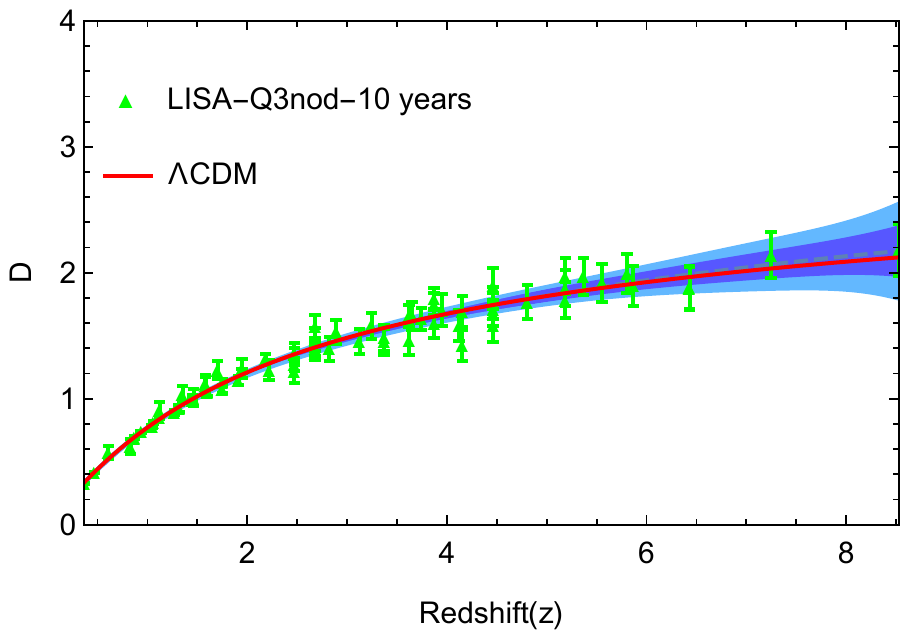}\\
\includegraphics[width=0.3\textwidth]{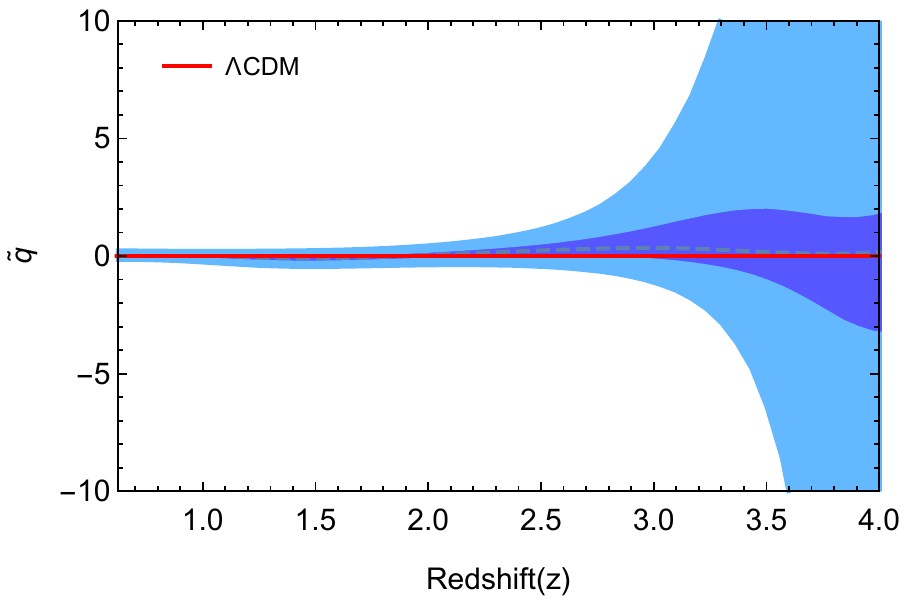}
\includegraphics[width=0.3\textwidth]{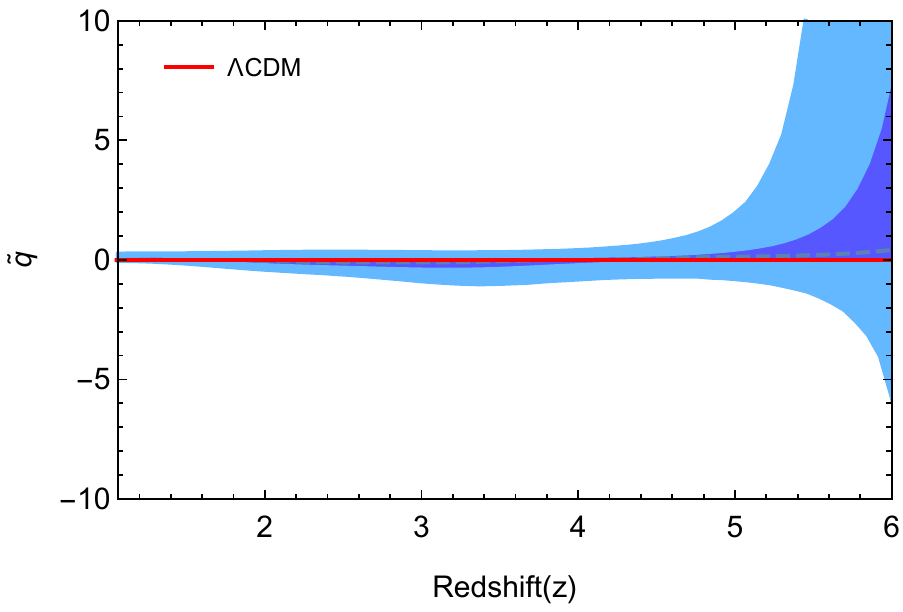}
\includegraphics[width=0.3\textwidth]{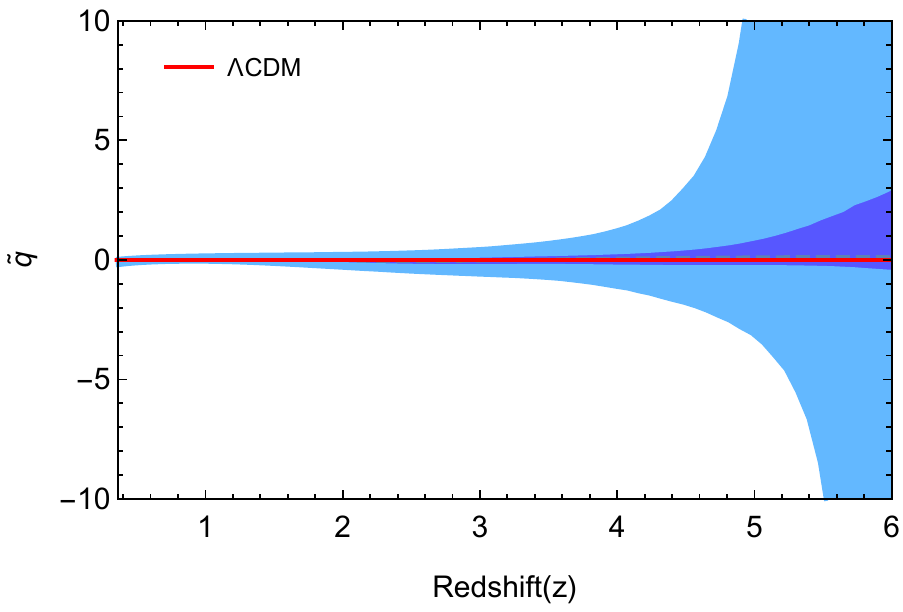}
\caption{Reconstructions of the interaction using LISA alone for a 10 years mission. From left
  to right each column reports the results for popIII, Q3d, Q3nod.}
\label{fig:dark_LISA_10y}       
\end{figure}

\begin{figure}[h]
\centering
\includegraphics[width=0.3\textwidth]{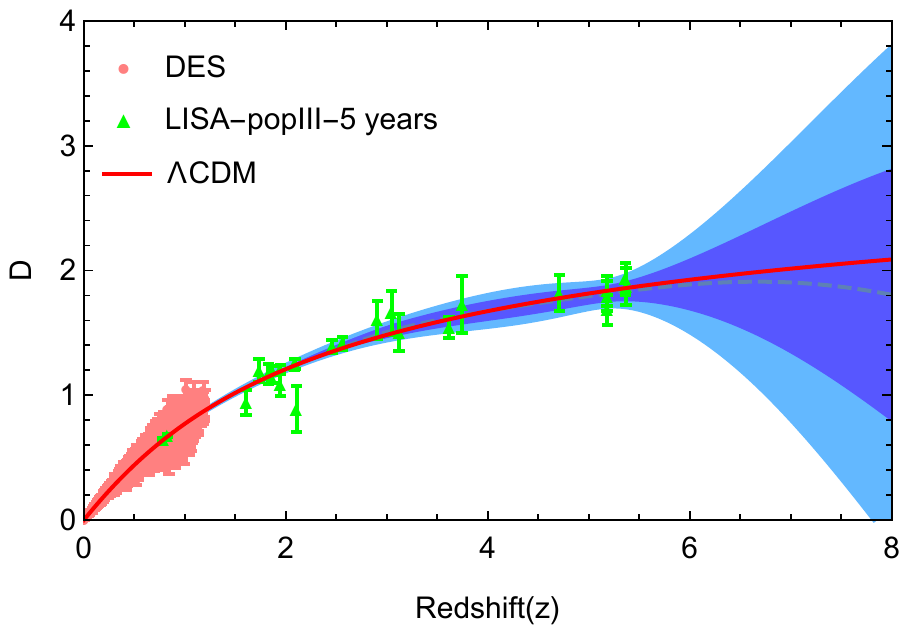}
\includegraphics[width=0.3\textwidth]{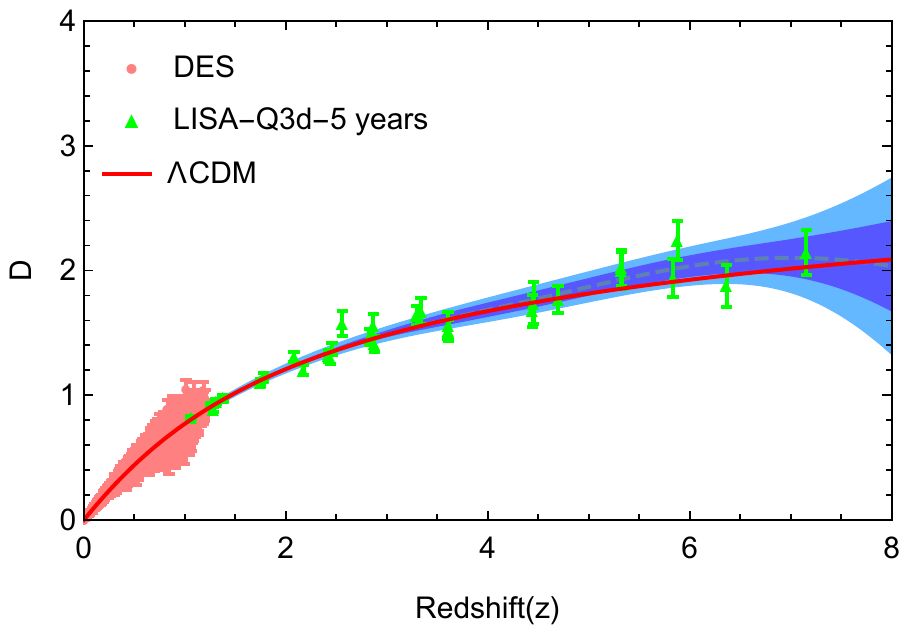}
\includegraphics[width=0.3\textwidth]{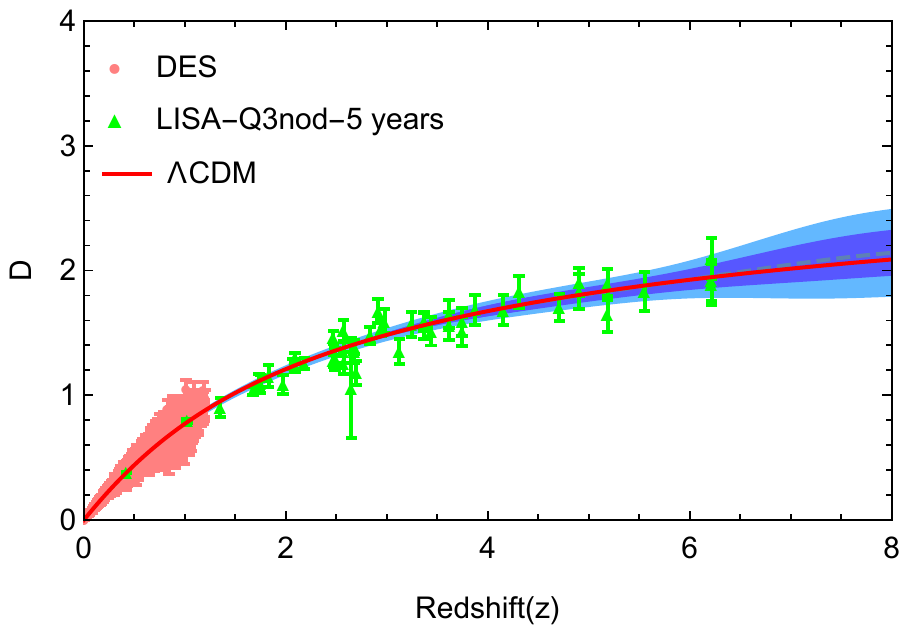}\\
\includegraphics[width=0.3\textwidth]{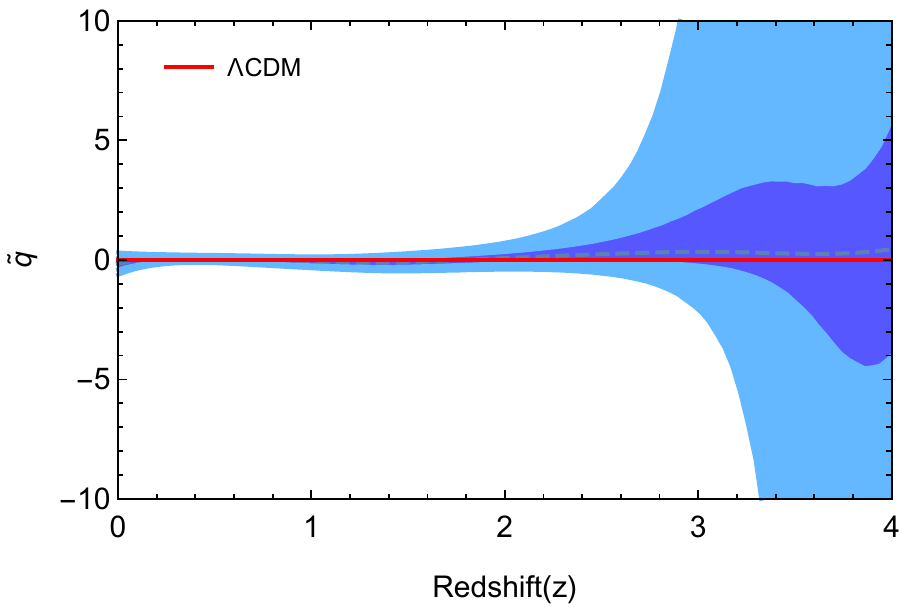}
\includegraphics[width=0.3\textwidth]{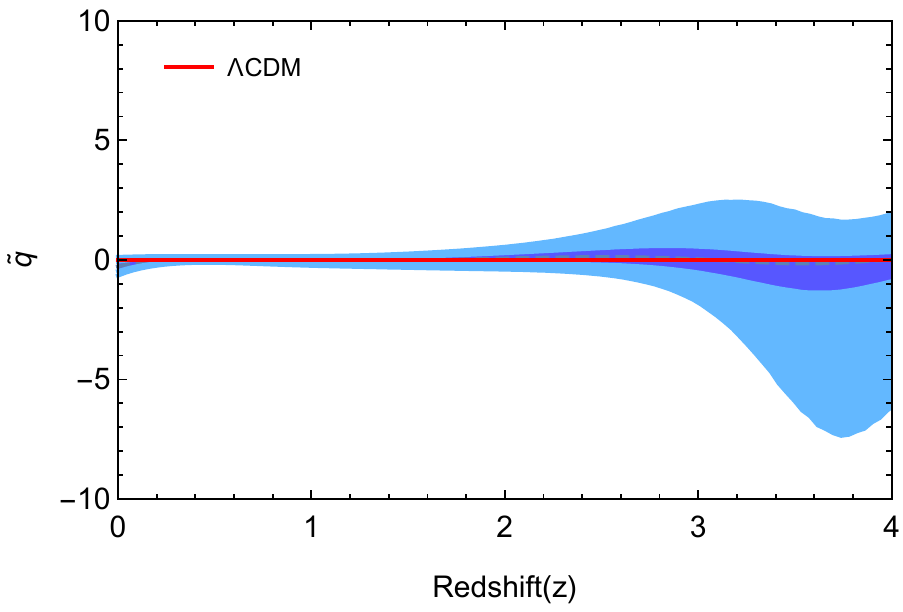}
\includegraphics[width=0.3\textwidth]{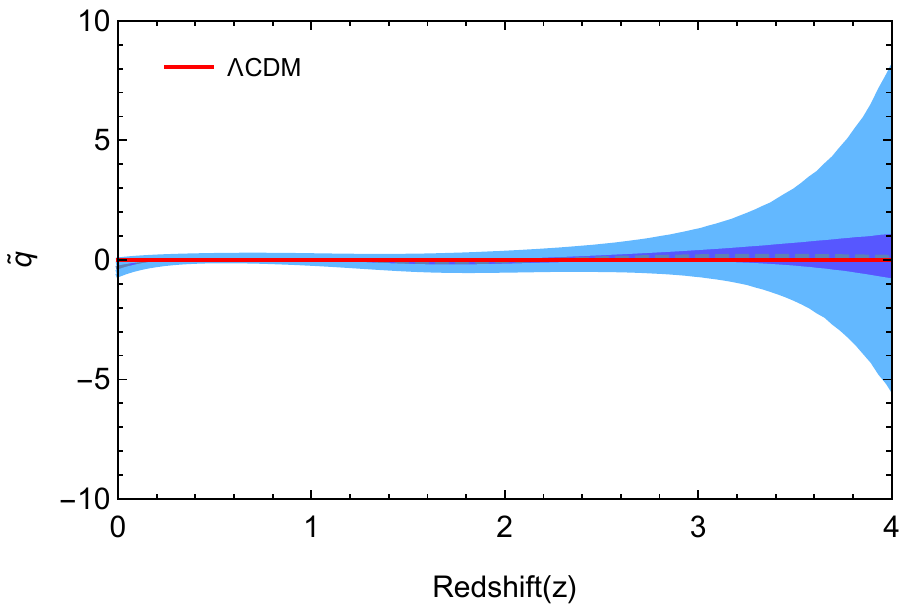}
\caption{Reconstruction of the interaction using DES$+$LISA (5 years) data. From left to right each column reports the results for popIII, Q3d, Q3nod.}
\label{fig:dark_LISA_5y_LpD}       
\end{figure}

\begin{figure}[h]
\centering
\includegraphics[width=0.3\textwidth]{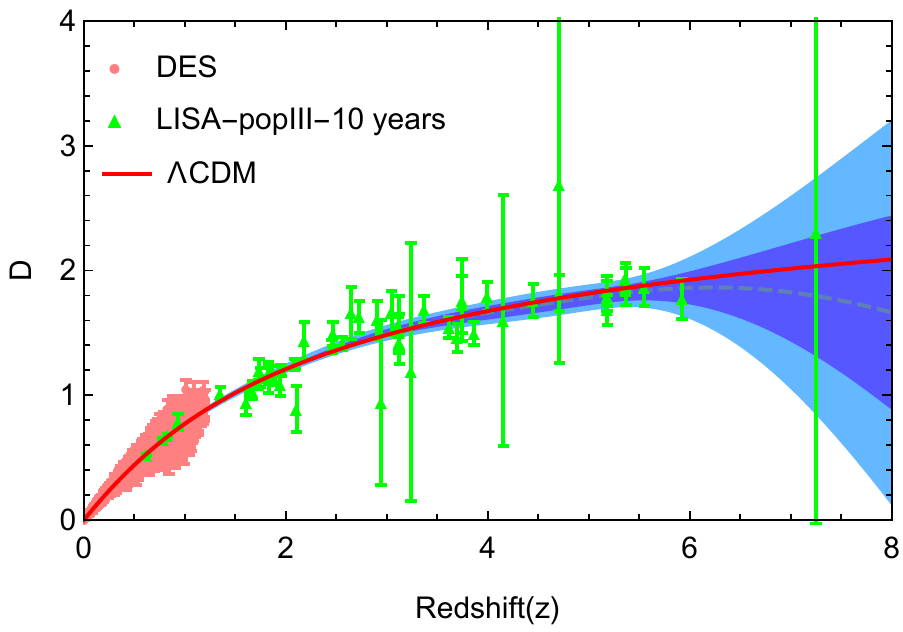}
\includegraphics[width=0.3\textwidth]{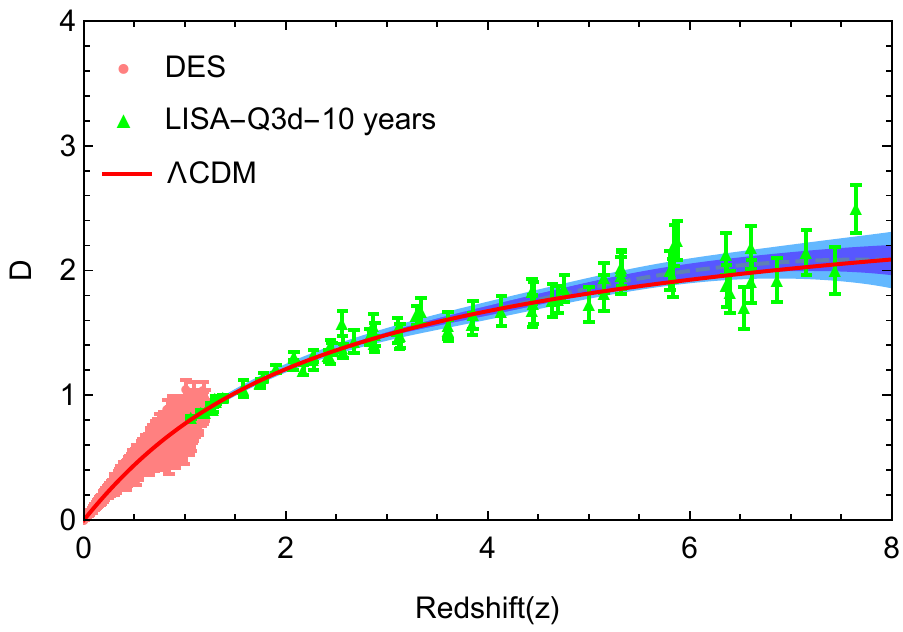}
\includegraphics[width=0.3\textwidth]{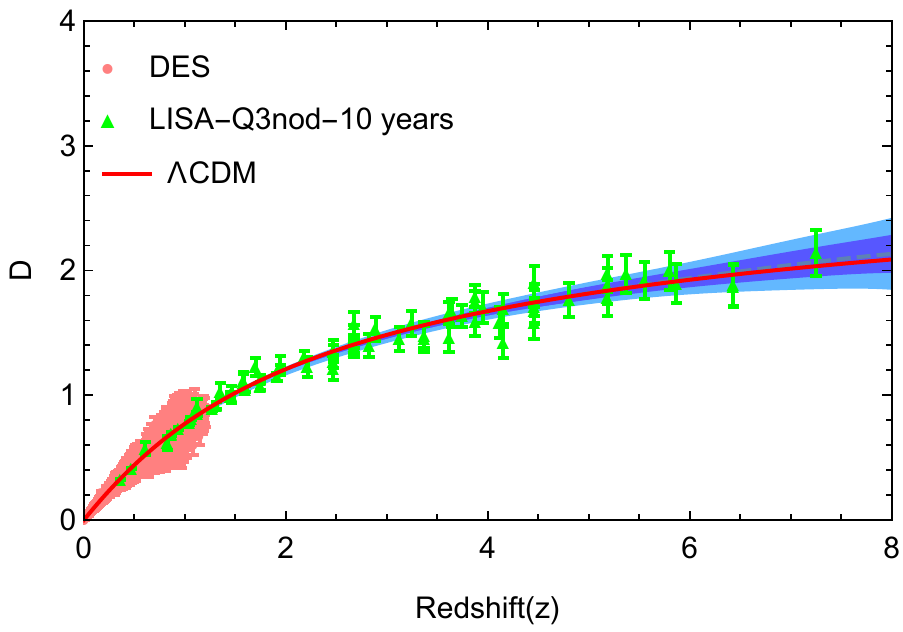}\\
\includegraphics[width=0.3\textwidth]{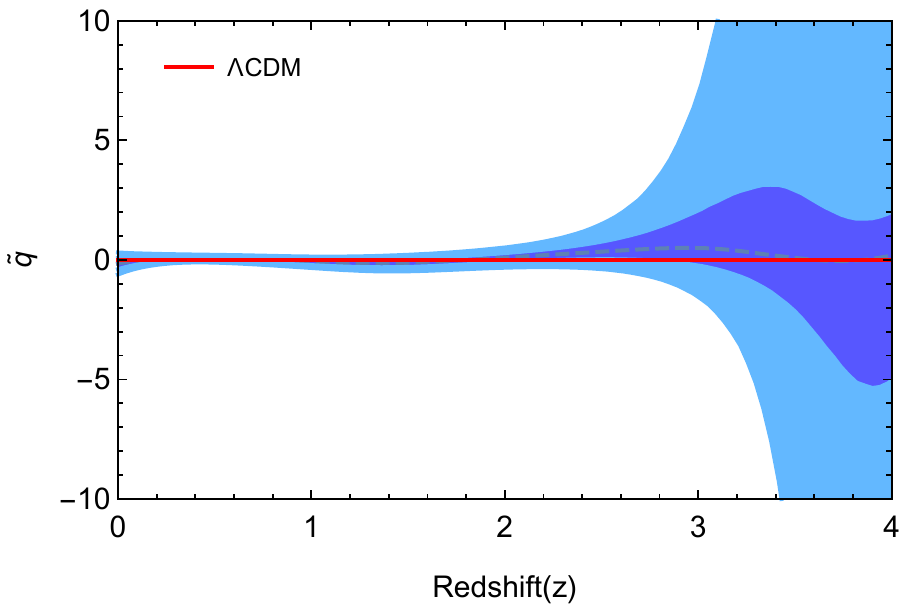}
\includegraphics[width=0.3\textwidth]{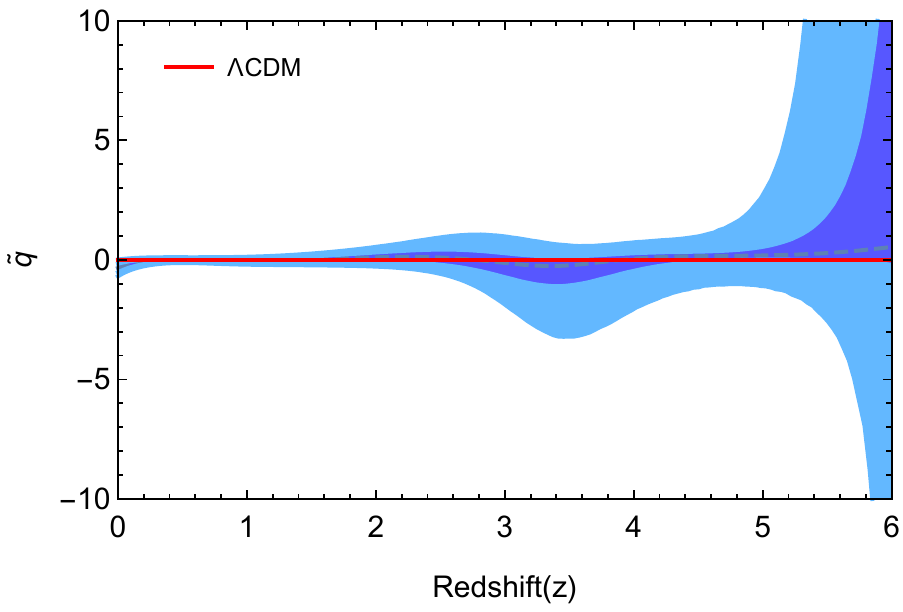}
\includegraphics[width=0.3\textwidth]{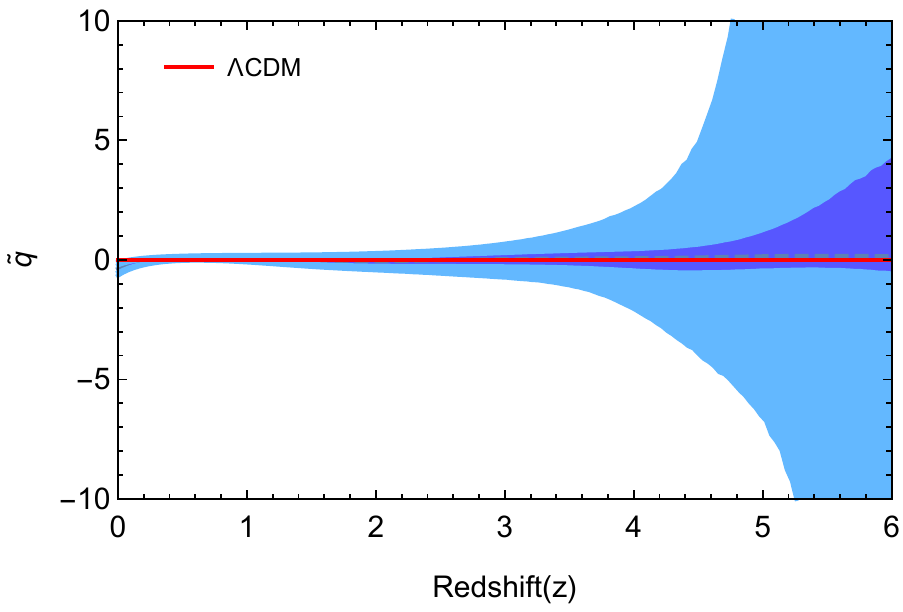}
\caption{Reconstruction of the interaction using DES$+$LISA (10 years) data. From left to right each column reports the results for popIII, Q3d, Q3nod.}
\label{fig:dark_LISA_10y_LpD}       
\end{figure}

We use the same methods developed in~\cite{Cai:2015zoa} for reconstructing the dark sector interaction with LISA. Our results are summarised by three different ``elements'': three initial conditions for the massive BH, two mission years, and two different data combinations (LISA alone or LISA $+$ DES).
All results from~\cite{Cai:2017yww} are shown in figures~\ref{fig:dark_LISA_5y}-\ref{fig:dark_LISA_10y_LpD}.
We can see that using MBHB standard siren alone, the simulated LISA data can reconstruct the interaction well from about $z\sim1$ to $z\sim3$ (for a 5 years mission) and $z\sim4$ or even $z\sim5$ (for a 10 years mission).
However the reconstruction is not efficient at redshift lower than $z\sim 1$ as MBHB events are much rare at late times.
When combined with the simulated DES datasets, Gaussian processes can reconstruct the interaction well in the whole redshift region from $z\sim0$ to $z\sim3$ for 5 years LISA and from $z\sim0$ to $z\sim5$ for 10 years LISA, respectively.
Thus the addition of DES data points, which cover the $0<z<1.5$ range, to the LISA MBHB standard siren catalogues, covering the range $1<z<10$, effectively reconstruct the dark interaction at all accessible redshift values, nicely integrating with each other.
We can also note that in all our results, the MBHB scenarios of Q3d and Q3nod are better than popIII at reconstructing the dark sector interaction, in agreement with the results of~\cite{Caprini:2016qxs}.

\begin{figure}[h]
\centering
\includegraphics[width=0.45\textwidth]{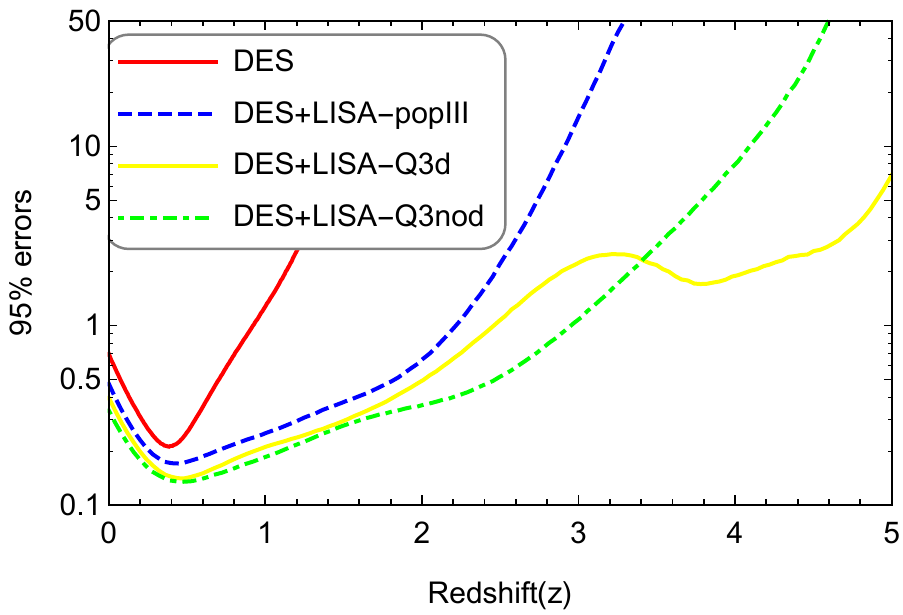}
\includegraphics[width=0.45\textwidth]{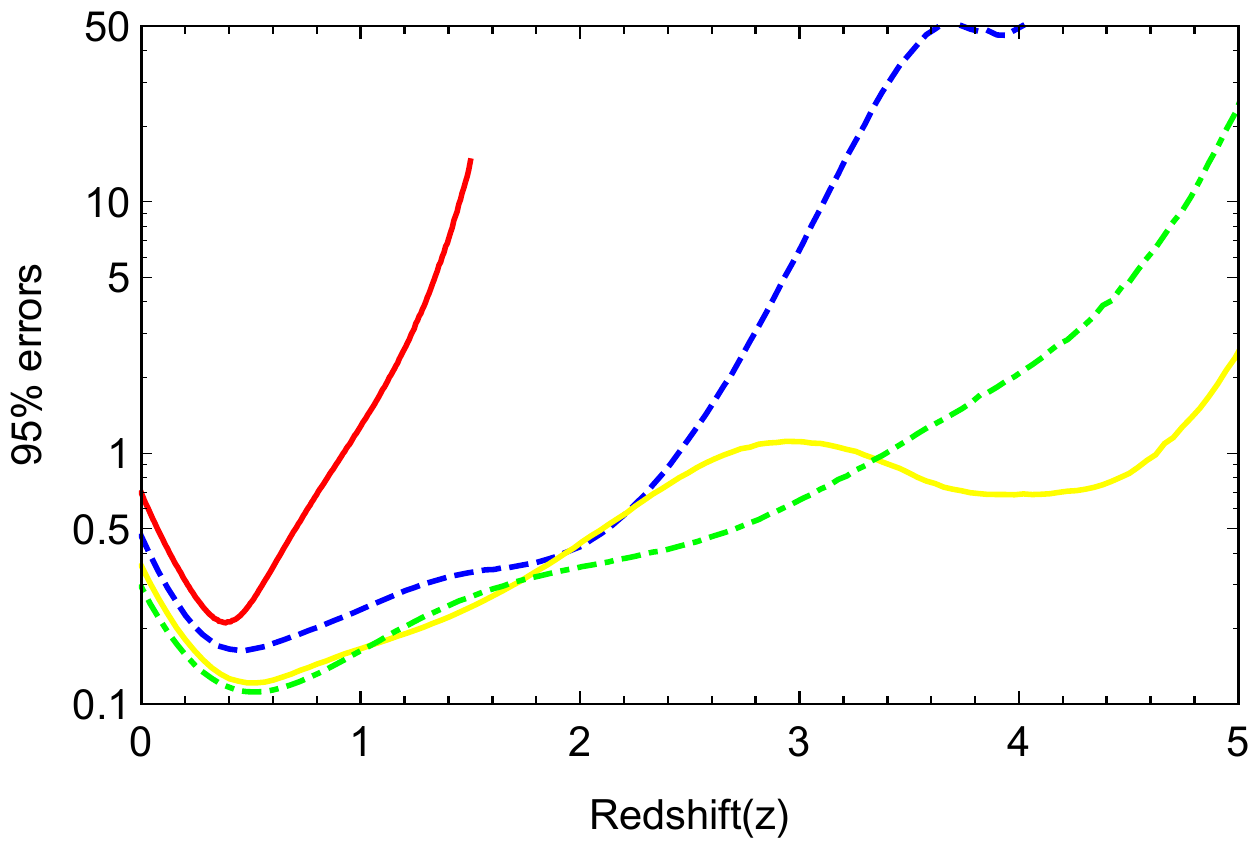}
\caption{95\% confidence errors on the reconstructed interaction term~\cite{Cai:2017yww}.}
\label{fig:err}       
\end{figure}

In order to understand how MBHB cosmological data will improve the constraints provided by traditional EM signal such as SNIa (standard candles), we can compare the reconstruction uncertainty provided by~\cite{Cai:2015zoa}, i.e.~using DES only data, with the ones obtained by DES+LISA data. The results are shown in figure~\ref{fig:err}.
We see that when the contribution of LISA
MBHB standard siren is added, the errors remain small up to $\sim3$ (for popIII) or $z\sim5$ (for Q3d), where they reach levels comparable to the value at $z\sim 1.5$ given by DES only.
Thus we can conclude that using only LISA data, the dark sector interaction can be
well reconstructed  from $z\sim 1$ to $z\sim 3$ (for a 5 years mission) and from $z\sim1$ to $z\sim5$ (for a 10 years mission). While, with DES datasets, the interaction is well reconstructed in
the whole redshift range from $z=0$ to $z\sim 3$ (5 years mission) and from $z=0$ to $z\sim5$ (10 years mission). Thus the massive BH binary can be used to constrain the dark
sector interaction at redshift ranges not reachable by the usual supernoave  datasets which probe only the $z <1.5$ range. So, LISA will extend the ability of SNIa data to constrain the
dark sector interaction in a model independent manner up to higher redshift.

\section{Constraint ability of GW on cosmological parameters with ET }

\begin{figure}[h]
\centering
\includegraphics[width=0.45\textwidth]{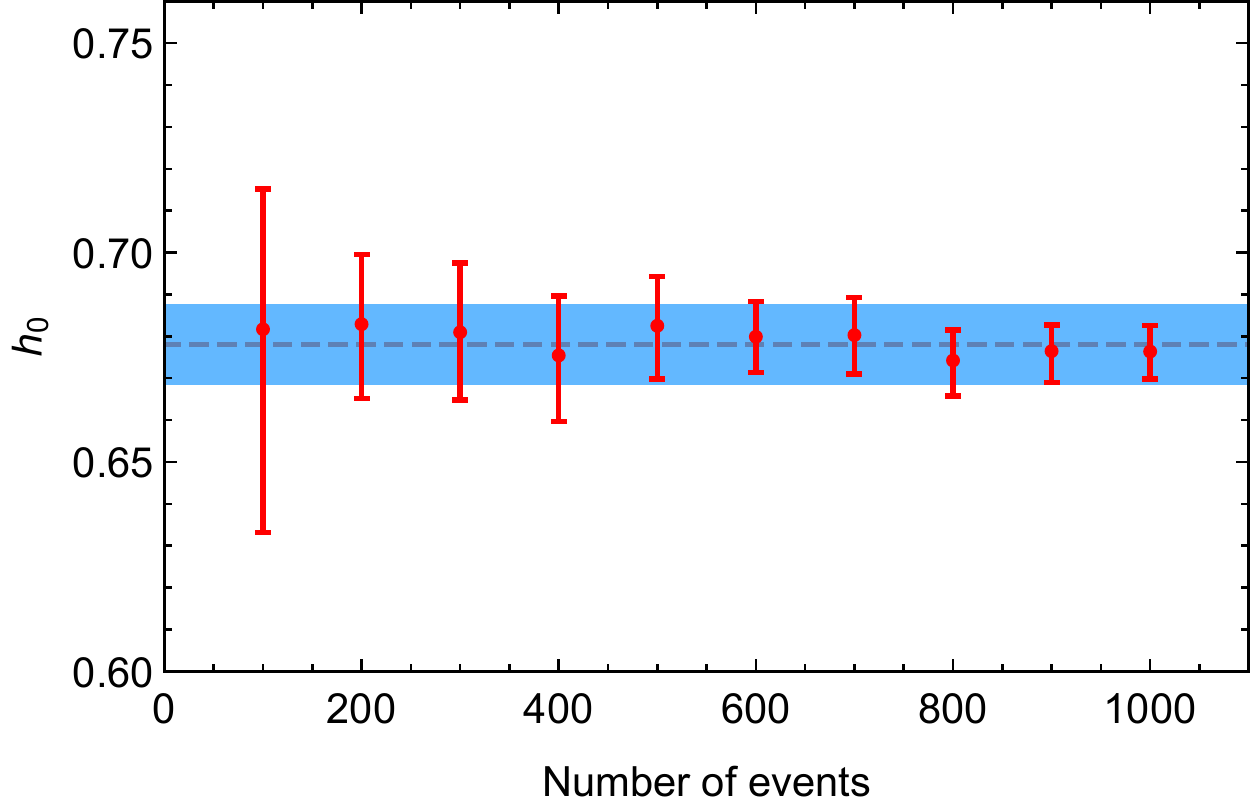}
\includegraphics[width=0.45\textwidth]{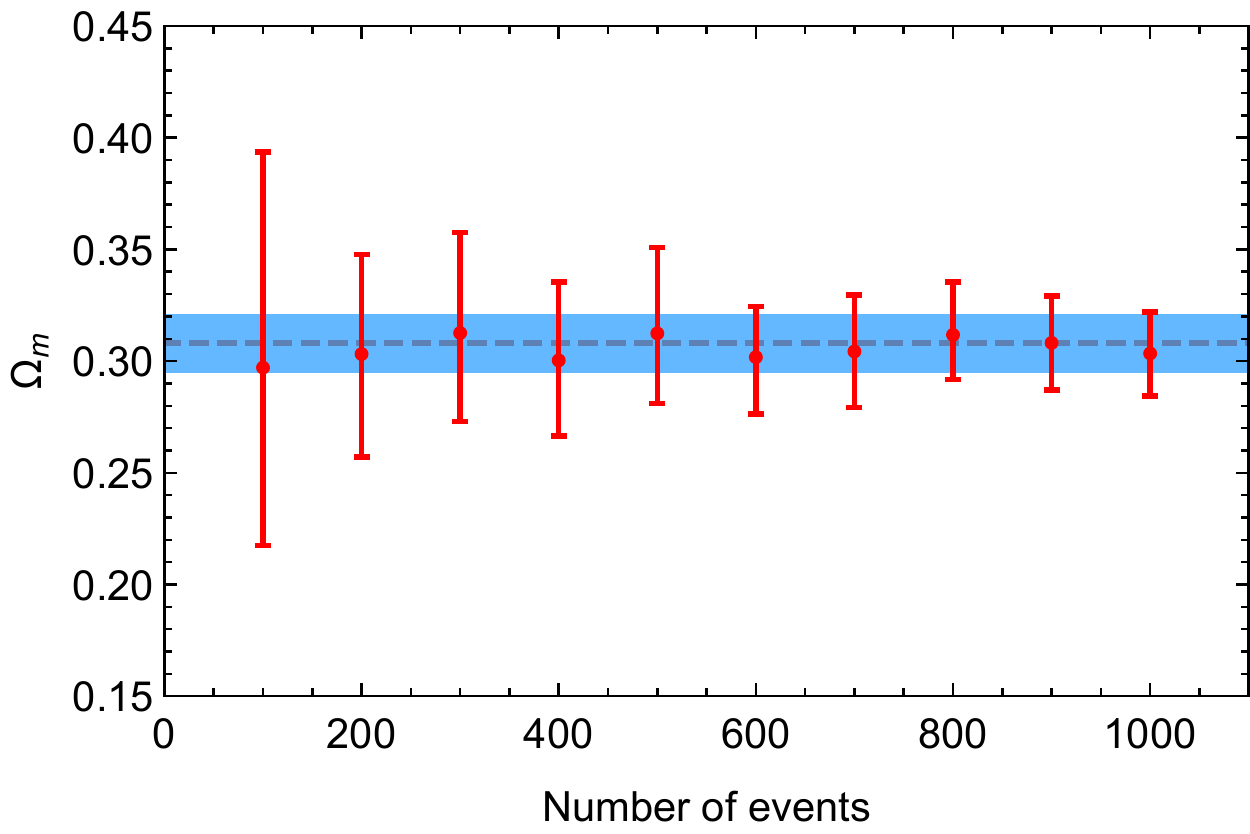}
\caption{Sixty-eight percent confidence level (C.L.) (red line) and the best fit (red dot) for
  $H_0$ (left) and $\Omega_m$ (right) for a
  variable number of GW events with EM counterpart. The fiducial model is shown as the dashed line. For a comparison, the
  blue shaded area is the $68\%$ C.L. constrained by the {\it Planck} temperature data combined with {\it Planck} lensing in the current
  {\it Planck} 2015 results.}
\label{fig:h_om}       
\end{figure}

This section briefly summarizes the results from the paper~\cite{Cai:2016sby} in which we use GWs as the standard sirens and check their abilities on constraining the cosmological parameters with ET. 
For ET, the GWs we mainly focus on are emitted by the binaries of NS-NS and NS-BH~\cite{Zhao:2010sz}. The binary merger of a NS with either a NS
or BH is hypothesized to be the progenitor of a short and intense burst of $\gamma$-rays (SGRB)~\cite{Nakar:2007yr}. An EM counterpart like SGRB can provide the redshift information if the host galaxy of the event can be pinpointed.
The expected rates of BNS and BHNS detections per year for the ET are about the order $10^3-10^7$ (ET project). However, only a small fraction ($\sim 10^{-3}$) is expected to have the observation of SGRB. In this work we take the detection rate in the middle
range ($10^5$), thus $10^2$ events with SGRB per year. With this postulate, we can simulate 100
to 1000 numbers of GW events to see that with how many events we can constrain the cosmological parameters as precisely as the current {\it Planck} results. We adopt the MCMC method and the results from~\cite{Cai:2016sby} are shown in figure~\ref{fig:h_om}.
We find that with about 500-600 GW events we can constrain the Hubble constant with an accuracy comparable to \textit{Planck} temperature data and \textit{Planck} lensing combined results. As for the dark matter density parameter, the GW data  alone seem not able to provide a constraint as good as for the Hubble constant, the sensitivity of 1000 GW events is a little lower than that of \textit{Planck} data. It should require more than 1000 events to match the \textit{Planck} sensitivity.

\begin{figure}[h]
\centering
\includegraphics[width=0.5\textwidth]{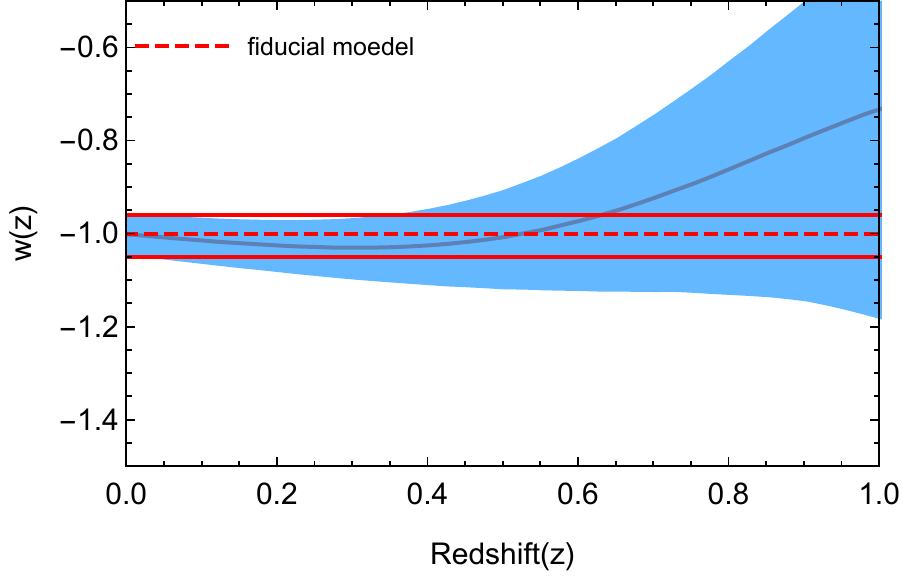}
\caption{Reconstructions of $w(z)$ from the simulated data sets with 700 GW events. The shaded blue regions are the $68\%$ C.L. of the constraint. The fiducial value $w=-1$ is shown as the red dashed line. For a comparison, we add the two red lines that bound the $68\%$ C.L. of the constraint of the constant equation of state $w$ from the \textit{Planck} data combined with the type-Ia supernovae. This figure is from~\cite{Cai:2016sby}.}
\label{fig:wplanck}       
\end{figure}

For constraining the equation of state $w(z)$, we adopt the nonparametric method Gaussian process to reconstruct the $w(z)$ function. We find 700 GW events can give the same constraint
accuracy to $w(z)$ as {\it Planck} (the constant w) in the low redshift region as shown in 
figure~\ref{fig:wplanck}. With 1000 GW events, we can constrain $\Delta h_0\sim 5\times 10^{-3}$ and $\Delta \Omega_m\sim 0.02$. The equation of state can be constrained to $\Delta w(z)\sim 0.03$ in the low redshift region with Gaussian process.

\section{Conclusions and future prospects }
GW standard sirens will in general offer a
new independent way, complementary to ordinary EM observations, to probe the cosmic expansion. Even if rare, GW standard sirens will complement, and increase confidence in other cosmic distance indicators, in particular SNIa standard candles. ET and LISA as the future GW detectors will play important roles in the GW astronomy. The different sources and the reshift regions supplied by the EM counterpart will enrich the information of the traditional EM signals. With the model-independent numerical method such as the Gaussian process, one can easily processes the GW's data sets and apply them to our researches. 

The machine learning as a big data analysis technique will become more and more important in the future. As one of the machine learning technique, Gaussian process can provide us a new and model-independent method to analyse the huge and complex data sets. It can help us to find the
physical meaning and forecast the tendency and property which are implied in these data sets.

Also, we can combine the standard sirens data with the {\it Planck} data and expect giving 
better constraints on the cosmological parameters. For example, we can use both the LISA and
ET simulated data sets combined with {\it Planck} data, then adopt the MCMC method to give the
constraints of a series of cosmological parameters. By adding the GWs data to the {\it Planck}
data, the tighter constraints of the parameters such as the $H_0$, $\Omega_m$, $w_0$, and $w_a$
are anticipated. 

\section{Acknowlodgements} 
We thank Z.K. Guo and Nicola Tamanini for helpful collaborations in those works discussed in this paper. This work was supported in part by  the National Natural Science Foundation of China Grants No.11690022, No.11375247, No.11435006, and No.11647601, and by the Strategic Priority Research Program of CAS Grant No.XDB23030100 and by the Key Research Program of Frontier Sciences of CAS.

\end{document}